\documentclass[preprint,longbib]{aastex701} 
\usepackage[utf8]{inputenc}
\usepackage{amsmath,amssymb}
\usepackage{booktabs}
\usepackage{amsthm}

\newcommand{\bmu}{\boldsymbol{\mu}}
\newcommand{\btheta}{\boldsymbol{\theta}}

\newcommand{\dmuk}{\Delta\boldsymbol{\mu}_k}
\newcommand{\bC}{\mathbf{C}}
\newcommand{\bD}{\mathbf{D}}
\newcommand{\bI}{\mathbf{I}}
\newcommand{\bQ}{\mathbf{Q}}
\newcommand{\bU}{\mathbf{U}}
\newcommand{\bV}{\mathbf{V}}
\newcommand{\bW}{\mathbf{W}}
\newcommand{\bL}{\mathbf{L}}
\newcommand{\br}{\boldsymbol{r}}
\newcommand{\ba}{\boldsymbol{a}}
\newcommand{\bb}{\boldsymbol{b}}
\newcommand{\bc}{\boldsymbol{c}}
\newcommand{\bg}{\boldsymbol{g}}

\newcommand{\bone}{\mathbf{1}}

\newcommand{\SNANA}{{\tt SNANA}}
\newcommand{\DESVYR}{{\tt DES-SN5YR}}

\newcommand{\chisq}{\chi^2}
\newcommand{\chisqmarg}{\chi^2_{\rm marg}}
\newcommand{\wwOm}{$w_0,w_a,\Omega_m$}

\newcommand{\covmat}{{\mathbf C}}   
\newcommand{\covsys}{{\mathbf C}_{\rm sys}}
\newcommand{\D}{{\mathbf D}}        
\newcommand{\rhoww}{\rho_{w_0w_a}}
\newcommand{\FoMratio}{R_{\rm FoM}}
\newcommand{\FoMWood}{{\rm FoM}_{\rm Woodbury}}
\newcommand{\FoMCinv}{{\rm FoM}_{{\covmat}^{-1}}}
\newcommand{\rk}[1]{\textcolor{brown}{#1}}

\newcommand{\ayan}[1]{\textcolor{blue}{#1}}

\newcommand{\URLcreatecov}{\url{https://github.com/RickKessler/SNANA/blob/master/util/create_covariance.py}}
\newcommand{\URLwfit}{\url{https://github.com/RickKessler/SNANA/blob/master/src/wfit.c}}
\newcommand{\URLRCC}{\url{https://rcc.uchicago.edu}}
\newcommand{\URLNERSC}{\url{https://www.nersc.gov/what-we-do/resources/perlmutter}}

\begin{document}
\title{Binning is Sinning: Absolution for the Unbinned Supernova Hubble Diagram}
\author[orcid=0000-0002-9436-8871,gname=Ayan,sname=Mitra]{Ayan Mitra}
\email[show]{ayan@illinois.edu}
\affiliation{Center for AstroPhysical Surveys, National Center for Supercomputing Applications, Urbana, IL 61801, USA}
\affiliation{Department of Astronomy, University of Illinois at Urbana-Champaign, Urbana, IL 61801, USA}

\author[orcid=0000-0002-4875-4157,gname=Anish,sname=Mukherjee]{Anish Mukherjee}
\email{anishm.space@gmail.com}
\affiliation{Department of Statistics, Rice University, Houston, TX, 77005, USA}

\author[orcid=0000-0002-4875-4157,gname=Richard,sname=Kessler]{Richard Kessler}
\email{kessler@kicp.uchicago.edu}
\affiliation{Kavli Institute for Cosmological Physics, University of Chicago, Chicago, IL 60637, USA}
\affiliation{Department of Astronomy and Astrophysics, University of Chicago, 5640 South Ellis Avenue, Chicago, IL 60637, USA}
\correspondingauthor{Ayan Mitra}

\begin{abstract}
Cosmological analyses of Type Ia supernovae use a covariance matrix to describe statistical noise and
systematic uncertainties.
To fit a Hubble diagram with an unbinned covariance matrix, the CPU cost scales as $O(N^3)$ for $N$ supernovae,
which has motivated compression into bins that loses information and reduces self-calibration power.
To remove the need for lossy compression,
we present a method that greatly reduces the CPU time for matrix inversion and cosmology fitting,
as well as the matrix storage.
For systematic covariances constructed from $K$ per-supernova distance-shift vectors, 
we retain the exact representation $\bC=\bD+\bU\bU^T$, with diagonal statistical covariance $\bD$,
and use the Woodbury identity to efficiently invert $\bC$. We use the Woodbury factorization identity to invert $\bC$,
which significantly reduces  computing time and disk  storage 
without losing information.
We implement this method in \SNANA\
and validate it on simulated samples for
LSST ($N{\approx}5{,}300$, $K=18$), 
Roman+LSST ($N{\approx}15{,}000$, $K=30$), and 
10-yr LSST ($N{\approx}134{,}400$, $K=18$).
For the 10-yr LSST case, the matrix inversion and cosmology fit time 
is ${\sim}10$~weeks using a dense (unbinned) matrix with 18 billion elements, 
and this time is reduced to ${\sim}1.5$~hr using the factorization method. 
The factorization method also reduces covariance matrix disk storage by 
a factor of 1,800 (36~GB to 20~MB).
In all validation tests, the best-fit cosmological parameters are nearly
identical to the unbinned fit, and the dark energy figure of merit (FoM) agrees to better than 0.1\% 
accuracy; the small differences are likely due to numerical artifacts in the pipeline.
For the LSST and Roman era, this factorization method enables 
large unbinned supernova analyses 
with significantly reduced computing resources compared to previous analyses using a dense matrix,
while preserving all information in the covariance model.
\end{abstract}

\section{Introduction}

A cosmological fit to Type Ia supernovae (SNe~Ia) compares observed distances versus redshift 
to a model prediction and minimizes a chi-square ($\chisq$) computed
from the data--model Hubble residuals and their covariance matrix ($\covmat$).
The matrix $\covmat$ has two parts: a diagonal part ($\D$) for each supernova's statistical measurement uncertainty, 
and a systematic part ($\covsys$) describing 
systematic uncertainties related to calibration, the SN~Ia light-curve model, Galactic extinction, etc.

Several past cosmology analyses \citep{Betoule2014,Scolnic2018,Brout2019} evaluated their $\chisq$ using a 
redshift-binned ($z$-binned) Hubble diagram (HD) and a $z$-binned $\covmat$
in order to reduce computation time. 
\citet{Brout2021} (BHS21) showed that $z$-binning results in weaker constraints
compared to fitting an unbinned Hubble diagram because the latter benefits from self-calibration
that reduces systematic uncertainties.

For spectroscopically confirmed SNe~Ia, using unbinned distances is straightforward.
For the photometrically identified sample used in the \DESVYR\ analysis \citep{Vincenzi2024},
\citet{Kessler2023} (K23) demonstrated how to determine unbinned distances using
the BEAMS with Bias Corrections \citep[BBC;][]{Kessler2017} likelihood for each event.
Although current SN samples  {with $N\sim 2{,}000$ events (where $N$ is the number of supernovae in the Hubble diagram)} are computationally feasible using the unbinned approach, 
K23 pointed out  that for the much larger samples expected from LSST and Roman, 
there will be severe CPU challenges for matrix inversion and 
$\chisq$ computation when $\covmat$ contains billions of elements.
To address this, K23 suggested a ``rebinned'' approach of binning the  {HD and $\covmat$} as a function
of redshift, color, and stretch, and showed that this approach recovers most, but not all,
of the  {constraining power} from an unbinned HD. 
More recent dark energy forecasts for LSST \citep[M26;][]{Mitra2025}
and Roman \citep[K25;][]{Kessler2025Roman} presented results using $z$-binned, rebinned, and unbinned distances,
confirming that while the rebinned method is significantly better than using $z$-binned distances,
it does not reach the full sensitivity of an unbinned HD because averaging events into discrete bins still washes out event-level information and degrades self-calibration.
Furthermore, the computing challenge for large (${\sim}10^4$) unbinned samples was already apparent in 
these forecasts: both studies used the unbinned method only for the nominal analysis
and relied on the rebinned method for the many systematic variants needed to quantify
individual systematic contributions.

Here we apply a matrix technique that greatly reduces the CPU and disk-storage requirements for 
processing large unbinned HDs, without losing information.
 {The key assumption is that} each systematic variation produces a vector of per-supernova distance shifts, 
and $\covsys$ is a sum of the outer products of $K$ such vectors.
For a fixed systematic model, $K$  {is independent of $N$.}
The key concept is to apply the Woodbury identity \citep{Woodbury1950} to $\covmat=\D+\bU\bU^T$, where $\bU$ collects the $K$ shift vectors. The inverse requires only a $K\times K$ matrix, replacing  {$O(N^3)$ unbinned matrix inversion and $O(N^2)$ per-point quadratic-form evaluations with a small-matrix setup and $O(N\times K)$ $\chisq$-likelihood evaluations} for $K\ll N$, while preserving the adopted covariance up to floating-point precision.

We have implemented this $\covmat$-factorization in the \SNANA\ program that builds 
$\covmat$,\footnote{\URLcreatecov}
updated the fast cosmology-fit code to use the factorized matrix,\footnote{\URLwfit}
and run validation tests across multiple sample sizes
to demonstrate the improvements in computing speed and storage.
Section~\ref{sec:formalism} presents the formalism,
and Section~\ref{sec:validation} describes the \SNANA\ implementation and validation tests
on three simulated samples: 
the LSST sample of M26,
the Roman+LSST sample of K25,
and an approximate 10-yr
LSST sample made by combining the 25 independent LSST realizations from M26.
Section~\ref{sec:storage} summarizes the storage scaling, and Section~\ref{sec:discussion} discusses broader implications before we conclude in Section~\ref{sec:conclusion}. Appendix~\ref{app:proofs} provides the mathematical proofs.

\section{Formalism}
\label{sec:formalism}

\subsection{Standard unbinned likelihood and marginalized chi-square}
\label{sec:marg_chi2}

We define $\bmu^{\rm obs}$ to be
an $N$-dimensional vector of observed distance moduli for $N$ supernovae, 
and $\bmu^{\rm model}(\btheta)$ to be
the corresponding vector of theoretical distance moduli as a function of cosmological model parameters $\btheta$.
For the spatially flat cosmological models considered in this work, $\btheta = (\Omega_m, w)$ for flat $w$CDM with a constant dark energy equation-of-state parameter $w$, or $\btheta = (\Omega_m, w_0, w_a)$ for flat $w_0w_a$CDM where the dark energy equation of state evolves with scale factor $a = 1/(1+z)$ as $w(a) = w_0 + w_a(1-a)$, and $\Omega_m$ is the present-day matter density parameter. The light-curve standardization coefficients (stretch $\alpha$ and color $\beta$) are determined prior to the cosmological fit when constructing the Hubble diagram.

The residual vector $\br(\btheta) = \bmu^{\mathrm{obs}} - \bmu^{\mathrm{model}}(\btheta)$ is a function of $\btheta$.
Following the standard approach \citep{Tripp1998,Guy2010,Conley2010}, a cosmological fit at fixed nuisance offset minimizes the quadratic form 
$\hat{\chisq}(\btheta) = \br(\btheta)^T {\covmat}^{-1} \br(\btheta)$, 
where $\covmat = \D + \covsys$ is the sum of the statistical ($\D$) and systematic ($\covsys$) covariance matrices.
Because supernovae measure only relative distances, the supernova absolute magnitude $M_B$ and Hubble constant $H_0$ enter $\br(\btheta)$ through a single degenerate additive offset that is marginalized analytically using a flat prior \citep{Conley2010}.
At each cosmology parameter point $\btheta$, the analytically marginalized chi-square is
\begin{equation}
    \chisqmarg(\btheta) \equiv \hat{\chisq}(\btheta) - \frac{B(\btheta)^2}{S}~,
    \label{eq:chi2marg}
\end{equation}
where
\begin{equation}  
    \hat{\chisq}(\btheta) = \br(\btheta)^T\bC^{-1}\br(\btheta),
    \qquad B(\btheta) = \bone^T\bC^{-1}\br(\btheta),
    \qquad S = \bone^T\bC^{-1}\bone.
   \label{eq:chi2marg_terms}
\end{equation}
Integrating over the additive offset with a flat prior gives, up to a parameter-independent constant, the marginalized log-likelihood
\begin{equation}
-2\log\mathcal{L}_{\rm marg}(\btheta) = \chisqmarg(\btheta) + \log|\bC| + \log S.
  \label{eq:marg_loglike}
\end{equation}
For the fixed covariances used in standard grid fits, both normalization terms $\log|\bC|$ and $\log S$ in Equation~\ref{eq:marg_loglike} are constant across parameter space, and therefore minimizing Equation~\ref{eq:chi2marg} or Equation~\ref{eq:marg_loglike} yields the same cosmological constraints.
In the conventional unbinned approach, forming and Cholesky-factoring the dense $N\times N$ matrix $\bC$ requires $O(N^3)$ computations (hence CPU time)
and $O(N^2)$ disk storage, 
and each evaluation of Equation~\ref{eq:chi2marg_terms} for a new trial parameter $\btheta$ requires $O(N^2)$
computations.
In the following subsection, we present the factorized approach that eliminates this bottleneck.

\subsection{Factorized covariance and efficient model fitting}
\label{sec:woodbury}

\begin{figure}[htbp]
\centering
\includegraphics[width=0.85\textwidth]{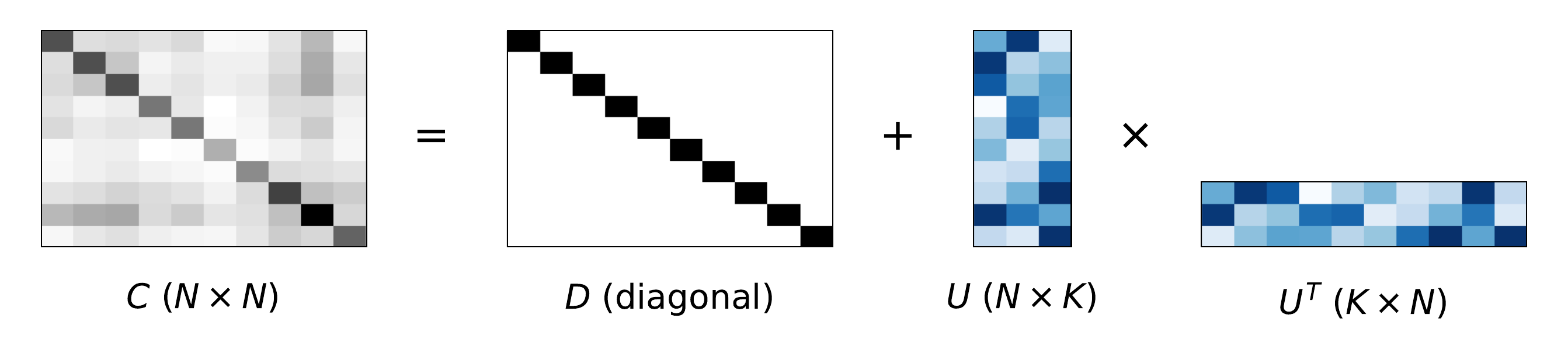}
\caption{Schematic of Equation~\ref{eq:cfull}: the covariance $\bC$ (left) is a diagonal matrix $\bD$ plus the product of a tall matrix $\bU$ (one column per systematic) and its transpose $\bU^T$. Shading is illustrative.}
\label{fig:schematic}
\end{figure}

Following Equation~8 of BHS21, the systematic covariance is constructed from $K$ per-supernova distance-shift vectors as
\begin{equation}
 \covsys = \sum_{k=1}^{K}\dmuk\,\dmuk^T = \bU\bU^T,
\label{eq:csyst}
\end{equation}
where $\dmuk$ is the $N$-dimensional vector of distance shifts of every supernova under the $k$-th systematic variation, already scaled by that systematic's assumed prior uncertainty, and $\bU=[\Delta\boldsymbol\mu_1\cdots\Delta\boldsymbol\mu_K]$ is the $N\times K$ matrix with one row per supernova and one column per systematic. The full covariance can therefore be written as
\begin{equation}
\covmat = \D + \bU\bU^T~,
\label{eq:cfull}
\end{equation}
a diagonal matrix plus a low-rank correction (Figure~\ref{fig:schematic}).
For a sample with $N=10^5$ events and $K=30$ systematics, $\covmat$ contains 
10 billion
entries,
whereas the factorized form $(\D, \bU)$ has only $N + N\times K = 3.1$~million elements without any loss of information.

Equation~\ref{eq:cfull} holds algebraically for any $N$ and $K$; numerical implementations agree up to floating-point round-off. This is not a truncation of an eigen-decomposition: $\bU$ is built directly from the  {same shift vectors used in the traditional $\covmat$-inversion approach.}
The Woodbury identity \citep{Woodbury1950,ShermanMorrison1950,Hager1989} and the matrix determinant lemma give
\begin{align}
\bC^{-1}&=\bD^{-1}-\bD^{-1}\bU\left(\bI+\bU^T\bD^{-1}\bU\right)^{-1}\bU^T\bD^{-1},\label{eq:woodbury}\\
\log|\bC|&=\log|\bD|+\log\left|\bI+\bU^T\bD^{-1}\bU\right|.\label{eq:logdet}
\end{align}
Only the $K\times K$ matrix $\bQ\equiv \bI+\bU^T\bD^{-1}\bU$ must be inverted (or Cholesky-factored), at a one-time setup cost of $O(NK^2+K^3)$. For a diagonal $\bD$ with positive entries and any real $N\times K$ matrix $\bU$, $\bQ$ is always symmetric positive definite and invertible.

Using Equation~\ref{eq:woodbury}, the marginalized chi-square quantities in Equation~\ref{eq:chi2marg_terms} are evaluated without ever forming an $N\times N$ matrix. 
We define $\ba = \bD^{-1} \br$, $\bb=\bU^T\ba$, $\bc=\bQ^{-1}\bb$, and $\bg=\bU^T\bD^{-1}\bone$, where $\ba$, $\bb$, and $\bc$ depends on $\btheta$.
Equation~\ref{eq:woodbury} gives
\begin{align}
\hat{\chisq}(\btheta) &= \br^T\ba-\bb^T\bc,\\
B(\btheta) & = \bone^T\ba-\bg^T\bc,\\
S & = \bone^T\bD^{-1}\bone-\bg^T\bQ^{-1}\bg.
\end{align}
Because $\bg$ does not depend on $\btheta$, $\bg$ and $S$ are computed once during setup. Each subsequent  {$\chisq$} evaluation for a new 
$\btheta$
costs $O(NK+K^2)$ operations, or $O(NK)$ for $K\ll N$, compared to $O(N^2)$ for an unbinned evaluation with a pre-factored dense covariance and $O(N^3)$ to factor it. To illustrate the scale of this reduction for the 10-yr
LSST sample ($N\approx 1.34\times 10^5$, $K=18$), the one-time setup drops from $N^3\sim 2.4\times 10^{15}$ to $NK^2\sim 4.3\times 10^7$ operations (nearly eight orders of magnitude fewer), and each $\chisq$ evaluation drops from $N^2\sim 1.8\times 10^{10}$ to $NK\sim 2.4\times 10^6$ operations (a factor of ${\sim}7{,}500$ reduction per grid point). Similarly, for the Roman+LSST sample ($N\approx 1.46\times 10^4$, $K=30$), setup drops from $N^3\sim 3.1\times 10^{12}$ to $NK^2\sim 1.3\times 10^7$ operations and per-point $\chisq$ evaluation drops from $N^2\sim 2.1\times 10^8$ to $NK\sim 4.4\times 10^5$ operations (nearly $500\times$ fewer).
Note that if $\bD$ or $\bU$ depends on fitted parameters, these setup terms must be recomputed whenever the covariance changes.

In standard supernova pipelines, systematic variations enter as independent uncorrelated effects so that the systematic weight matrix is the identity ($\bW=\bI$). More generally, if correlated systematic amplitudes with a $K\times K$ covariance matrix $\bW=(w_{kk'})$ are present (e.g. peculiar velocity covariances), 
such that $\covsys = \sum_{k=1}^{K}\sum_{k'=1}^{K} w_{kk'}\,\dmuk\,\Delta\boldsymbol{\mu}_{k'}^T = \bU\bW\bU^T$, one can compute the $K\times K$ Cholesky factor $\bW=\bL\bL^T$ and replace $\bU$ by $\bU'\equiv\bU\bL$; this restores $\covmat=\bD+\bU'\bU'^T$ and Equations~\ref{eq:woodbury}--\ref{eq:logdet} without requiring a separate matrix inversion of $\bW$.
An external positive-semidefinite covariance can be represented by multiple factor columns, although its rank may be large and remove the computational advantage. The current implementation exports the shift-vector contributions it constructs internally; if an additional external covariance is present, it skips the factorized output and warns. Non-diagonal statistical noise requires a different efficiently solvable base covariance. Our implementation works under the assumption of uncorrelated statistical noise, hence $\bD$ is diagonal.
Exactness here means equivalence to the adopted covariance model, rather than a claim that this model captures every physical source of uncertainty.
\section{Validation on Simulated Samples}

\subsection{Implementation in SNANA}
\label{subsec:snana}
With the new factorization code implemented in {\tt SNANA}, the covariance program writes the diagonal statistical variance vector $\bD$ and the $N\times K$ shift-vector matrix $\bU$ directly from the per-systematic shift vectors, and  {does not construct, invert, or write} the $N\times N$ matrix $\covmat$.

The cosmology-fitting program reads $\bD$ and $\bU$, 
and before beginning the fit it
precomputes the $K\times K$ matrix $\bQ=\bI+\bU^T\bD^{-1}\bU$, its Cholesky decomposition, the vector $\bg=\bU^T\bD^{-1}\bone$, and the scalar $S$. 
At each step of the minimization, it 
evaluates Equations~\ref{eq:woodbury} and \ref{eq:logdet} in place of a double loop over $\covmat^{-1}$.
The analytic marginalization over the distance-modulus offset, prior evaluations, and all other parts of the fit remain unchanged, and the fitting program never allocates or forms an $N\times N$ matrix in memory.

\newcommand{\fnMarg}{More efficient samplers can significantly reduce the fit time 
compared to our illustrative brute-force marginalization method. However, CPU fit time ratios 
are valid regardless of minimization technique.}

\subsection{Validation Tests}
\label{sec:validation}
We test the factorization on three simulated samples from the literature that have been used to forecast cosmology constraints:
(1) the LSST Deep Drilling Field sample of \citet{Mitra2025} ({M26}), 
with $N\approx5{,}300$ supernovae and $K=18$ systematics in each of 25 independent realizations;
(2) the Roman+LSST sample of \citet{Kessler2025Roman} (K25), with $N\approx14{,}600$ and $K=30$ in each of 
\rk{9} realizations; and
(3) an anticipated 10-yr
LSST sample with $N{\approx}134{,}400$  and $K=18$, 
made by combining the 25 realizations of M26.
The forecast quantity is the dark energy figure of merit (FoM) from a cosmology fit using the 
$w_0w_a$CDM model with a \wwOm\ marginalization\footnote{\fnMarg} 
grid size of $91^3$ for M26 and 10-yr LSST, 
and $81^2\times 37$ grid size for K25.
For consistent CPU comparisons, our reported  fit times for K25 have been scaled by 
$(91^3)/(81^2{\times}37) = 3.1$.    
Following the Dark Energy Task Force \citep[DETF;][]{Albrecht2006}, 
these forecasts use
the FoM definition of \citet{Wang2008}, $\mathrm{FoM}=[\sigma_{w_0}\sigma_{w_a}\sqrt{1-\rhoww^2}]^{-1}$, 
where $\rhoww$ is the fitted parameter correlation. 
To quantify the impact of factorization,
we re-ran the K25 and M26 covariance construction and cosmology fitting using
the \SNANA\ factorization updates (\S\ref{subsec:snana}), 
and evaluated the average figure of merit, $\FoMWood$.
Using the published $\FoMCinv$ based on the traditional unbinned $\covmat$,
we define the FoM ratio
\begin{equation}
    \FoMratio \equiv \frac{\FoMWood}{\FoMCinv}~,
    \label{eq:fom_ratio}
\end{equation}
and define  $\FoMratio - 1$ 
to be the numerical accuracy of the factorization approach.
Table~\ref{tab:summary} summarizes the 
CPU processing time, 
hard drive storage, 
and accuracy comparisons across all three samples. 
Across the validation samples, $|\FoMratio - 1| < 10^{-3}$, and
the small non-zero residual in K25 is from file-output rounding and grid-boundary interpolation artifacts.
The covariance construction speed is accelerated by a factor of ${\sim}10$, 
disk storage is reduced by two to three orders of magnitude,
and the cosmology fit speed is increased by factors of 6 and 60 for 
the M26 and K25 samples, respectively.

For the 10-yr LSST sample, we did not run the analysis using the unbinned method 
because of the large compute resources needed. However, we can very roughly
estimate the CPU time by assuming that the $\covmat$-inversion time scales as $N^3$
and the cosmology fit time scale as $N^2$; the result is ${\sim}1$~week to compute
$\covmat^{-1}$, and ${\sim}9$~weeks to perform the cosmology fit.

\newcommand{\fnCluster}{K25 was run on the Midway2 cluster at \URLRCC;
 M26 ran on Perlmutter at {\URLNERSC}
 }
\newcommand{\fnCsizeOld}{For unbinned, a {\tt Numpy} table stores the upper-triangular matrix in $32$-bit precision; for factorized, a {\tt Numpy} table stores the length-$N$ diagonal $\bD$ and $N\times K$ matrix $\bU$ in $64$-bit precision.
}
\newcommand{\fnCsize}{Stored in compressed {\tt Numpy} table. Unbinned $\covmat$ includes
upper-triangular matrix with 32-bit precision; factorized $\covmat$ includes 
diagonal $\bD$ and $N\times K$ matrix $\bU$ with $64$-bit precision.
}
\newcommand{\fnRebin}{Size of HD rebinning for \{redshift, colour, stretch\} is 
$14 \times 8 \times 4 =448$ for M26, and $45 \times 8 \times 4 =1440$  for K25.
}
\newcommand{\fnCPUnote}{CPU time for  stat+syst covariance matrix}

\newcommand{\fnLSST}{Mean over 25 realizations: 
$N\approx 5{,}300$, $K=18$,  \wwOm\ cosmology fit grid size = $91^3$.
}
\newcommand{\fnRoman}{
Mean over 9 realizations: $N\approx 14{,}600$, $K=30$, original \wwOm\ cosmology fit grid size = $81^2{\times}37$,
hence fit CPU times are scaled by $3.1$ to  compare with the $91^3$ grid size for M26).
}
\newcommand{\fnTENyrLSST}{
Combines the 25 realizations from M26: $N=134{,}387$, $K=18$, \wwOm\ cosmology fit grid size = $91^3$.
}
\newcommand{\fnCfact}{
Combines the 25 factorized $(\bD,\bU)$ from M26.
}
\begin{table}[htbp]
\centering
\renewcommand{\thempfootnote}{\arabic{mpfootnote}}
\caption{Summary of the validation tests
for covariance matrix ($\covmat$) construction and cosmology fitting. 
The performance metrics are CPU, disk storage, and 
$\FoMratio-1$ (Eq.~\ref{eq:fom_ratio}).
 }
\label{tab:summary}
\footnotesize
\setlength{\tabcolsep}{3.5pt}
\begin{tabular}{@{}lccccccccc@{}}
\toprule
 & \multicolumn{6}{c}{CPU\footnote{\fnCluster} time (minutes) for} & \multicolumn{2}{c}{$\covmat$ Storage\footnote{\fnCsize} } & \\
\cmidrule(lr){2-7}
 & \multicolumn{3}{c}{Covariance construction\footnote{\fnCPUnote} } & \multicolumn{3}{c}{Cosmology fit} & 
       \multicolumn{2}{c}{size (MB) } & \\
\cmidrule(lr){2-4}\cmidrule(lr){5-7}\cmidrule(lr){8-9} 
Sample & Unbin. & Fact. & Rebin.\footnote{\fnRebin}  & Unbin. & Fact. & Rebin. & Unbin. & Fact. &  {$\FoMratio - 1$} \\
\midrule
LSST (M26)\footnote{\fnLSST} & 1.8 & 0.2 & 0.2 & 145 & 25 & 14 & 57 & 1  & 0 \\
    Roman+LSST (K25)\footnote{\fnRoman} & 16 & 1.4 & 0.8 & 2,850 & 45 & 111 &  {425}  & 4  & 0.0008 \\
10-yr LSST\footnote{\fnTENyrLSST} & N/A\footnote{Not available because this test was not run.} & 6\footnote{\fnCfact} & N/A & N/A & 79 & N/A &  {36,120}  & 20 &  {N/A} \\
\bottomrule
\end{tabular} 
\end{table}

Across all three datasets, the factorized representation preserves the unbinned covariance and cosmological constraints without loss of information: in all 25 M26 realizations, factorized versus unbinned $\chisq$ evaluations agree to better than $10^{-15}$ prior to file-output rounding. In the cosmological fits, the best-fit central values and uncertainties on $w_0$, $w_a$, and $\Omega_m$, and the minimum $\chisq$ from the factorized covariance are identical to those from the unbinned covariance in every realization of the M26 and K25 samples. For M26, $\FoMratio-1 = 0$ across all 25 realizations (mean $\mathrm{FoM}=155.5\pm 2.8$ in both), and for K25, $|\FoMratio-1| \le 8\times10^{-4}$ ($0.08\%$, differing by at most $10^{-4}$ in $\sigma_{w_a}$ in three of nine realizations and $0.4$ in FoM in two of nine, with exact agreement in all stat-only fits).

By contrast, when using the rebinned covariance, the fitted parameters shift relative to the unbinned analysis: 
the (rebinned $-$ unbinned) shifts (quoting the mean shift and its standard error $\mathrm{std}/\sqrt{N_{\rm real}}$), for M26 are  $\Delta w_0 = +0.027 \pm 0.005$ 
and $\Delta w_a = -0.063 \pm 0.030$, 
with a mean FoM shift  $=\ayan{+}3.1 \pm 2.8\%$ (mean $\mathrm{FoM}=158.9\pm 2.5$, 
ranging from $-16\%$ to $+46\%$ across realizations). For 
K25, the ensemble shifts are $\Delta w_0 = -0.001 \pm 0.005$ (up to $0.02$) and $\Delta w_a = +0.017 \pm 0.017$ (up to $0.11$), and mean FoM shift  $= -5.5 \pm 0.9\%$ (mean $\mathrm{FoM}=479.2\pm 10.5$ versus $507.3\pm 11.1$, ranging from $-1\%$ to $-9\%$). Replacing the rebinned approximation with the factorized unbinned covariance eliminates these binning shifts while retaining fast runtimes. Because factorized unbinned inference matches the computational speed and compact storage of rebinning without losing information or biasing cosmological parameters and the dark energy FoM, we recommend discontinuing rebinned Hubble-diagram 
analyses\footnote{The rebin method has been used only in forecasts from simulations, not for real data.} 
and using
factorized unbinned inference for future supernova cosmology pipelines.

As summarized in Table~\ref{tab:summary}, the computational 
speedup grows rapidly with sample size. 
For M26, building a factorized covariance matrix is nearly ${\times}10$
faster than building and inverting the corresponding unbinned matrix ($0.2$~min versus $1.8$~min per matrix), 
and the $w_0w_a$CDM cosmology fit is nearly ${\times}6$
faster. 
For the larger K25 sample,
covariance construction is $11\times$ faster and the $w_0w_a$CDM cosmology fit
is $63\times$ faster than the unbinned fit, and $2.5\times$ faster than the rebinned fit. 
At full ten-year LSST scale ($N=134{,}387$, $K=18$), where building and fitting the $36$~GB 
unbinned covariance would require months of CPU time, 
the unbinned factorized approach completes in 79 minutes
using a $20.4$~MB covariance file.

\begin{figure}[tbp]
\centering
\includegraphics[width=0.82\textwidth]{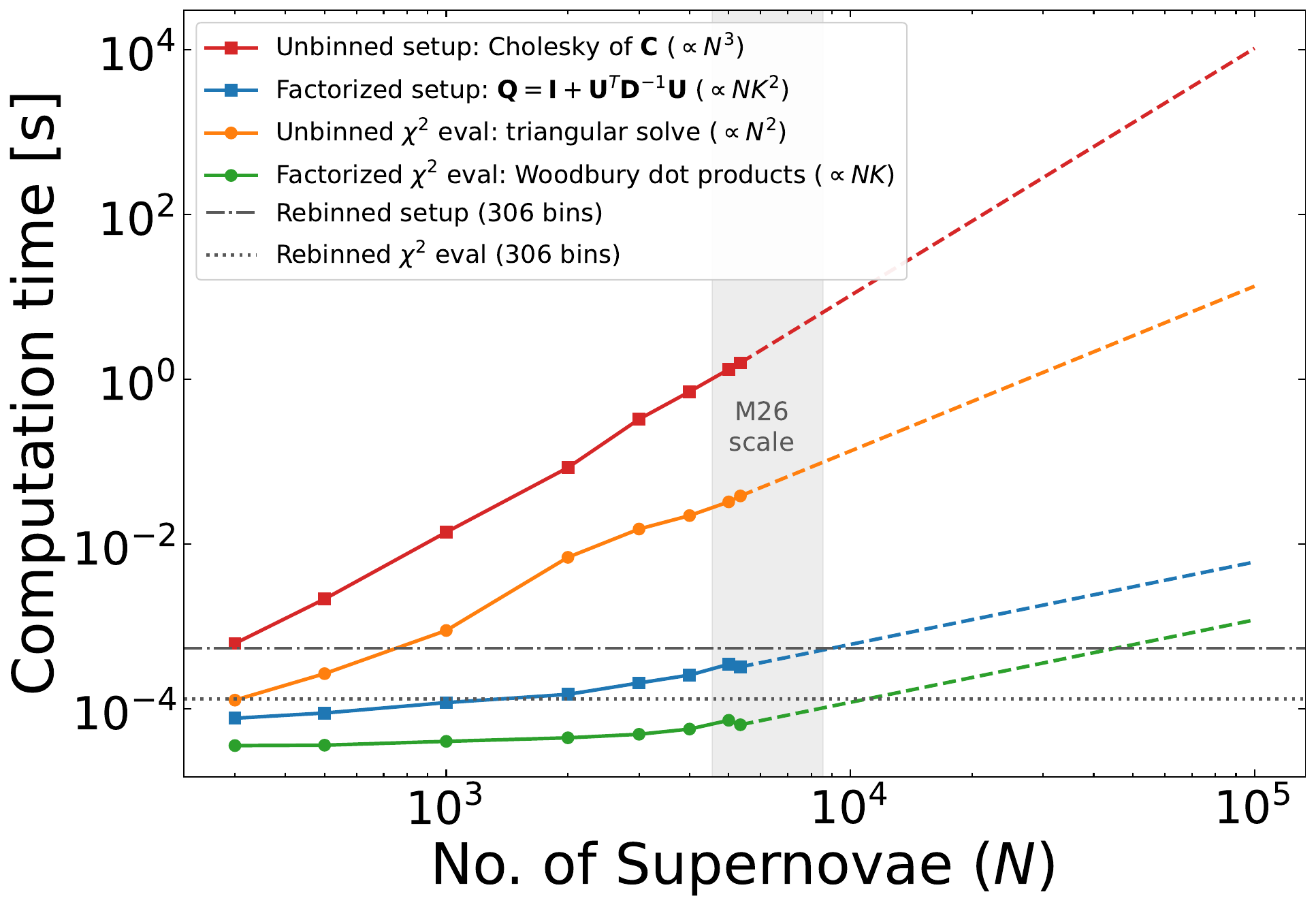}
\caption{One-time linear-algebra setup time (squares; $O(N^3)$ Cholesky factorization of $\bC$ for unbinned versus $O(NK^2)$ formation and Cholesky factorization of the $K\times K$ matrix $\bQ=\bI+\bU^T\bD^{-1}\bU$ for factorized) and per-point $\chisq$ likelihood evaluation time (circles; $O(N^2)$ triangular solve versus $O(NK)$ Woodbury dot products) as a function of the number of supernovae $N$ at fixed $K=18$. Solid curves are measured on M26 subsets up to $N=5{,}344$ (shaded band); dashed lines show the asymptotic scaling extrapolations ($N^3$, $N^2$, and $N$), and horizontal lines show the constant 306-bin rebinned setup and evaluation costs.}
\label{fig:nscaling}
\end{figure}

Figure~\ref{fig:nscaling} shows the one-time setup time and per-point $\chisq$ evaluation time as a function of $N$ at fixed $K=18$, for M26  and extrapolated to $N=10^5$. At $N=5{,}344$, factorized evaluation is ${\sim}600\times$ faster than an unbinned triangular solve ($0.064$~ms versus $38.4$~ms) and setup is ${\sim}4{,}900\times$ faster than unbinned Cholesky decomposition, growing to projected speedups of ${\sim}10^4\times$ and ${\sim}10^6\times$, respectively, at $N=10^5$.

\begin{figure}[htbp]
\centering
\includegraphics[width=0.82\textwidth]{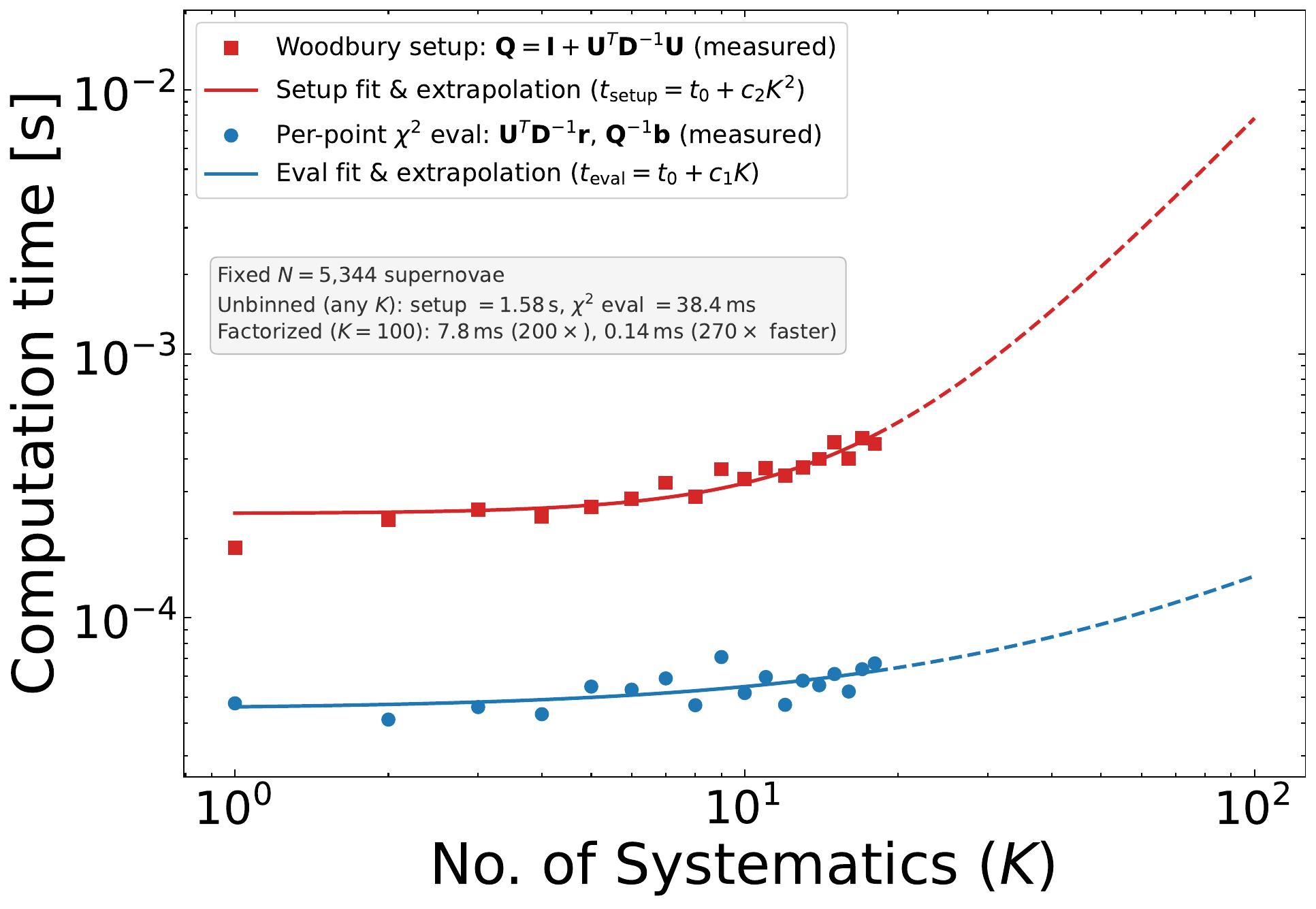}
\caption{Woodbury linear-algebra setup time (red squares; forming and factoring the $K\times K$ matrix $\bQ=\bI+\bU^T\bD^{-1}\bU$) and per-point $\chisq$ likelihood evaluation time (blue circles; evaluating $\bb=\bU^T\bD^{-1}\br$ and $\bc=\bQ^{-1}\bb$) as a function of the number of systematics $K$ at fixed $N=5{,}344$. Points are measured for $K=1$ to $18$ (minimum of 11 repeats, each averaging $2{,}000$ evaluations); curves show fits to $t_{\rm setup}(K)=t_0+c_2 K^2$ and $t_{\rm eval}(K)=t_0+c_1 K$ over the measured range (solid) and extrapolated to $K=100$ (dashed). The $K$-independent $N\times N$ unbinned reference times at $N=5{,}344$ are indicated on the plot.}
\label{fig:kscaling}
\end{figure}

Figure~\ref{fig:kscaling}
shows the complementary dependence on the number of systematics $K$ at fixed $N=5{,}344$, obtained by truncating the shift-vector matrix $\bU$ to its first $K$ columns. 
In the regime $K\ll N$, the Woodbury setup cost
grows quadratically with $K$, while the per-point likelihood evaluation cost
grows linearly with $K$. Because fixed $O(N)$ vector-operation overhead ($\ba=\bD^{-1}\br$ and $\br^T\ba$) dominates at small $K$, fitting the measured timings across $K=1\text{--}18$ to explicit overhead-plus-scaling models, $t_{\rm setup}(K) = t_{0,\rm setup} + c_2 K^2$ ($t_{0,\rm setup}\approx 0.25$~ms, $c_2\approx 0.76\,\mu\mathrm{s}$) and $t_{\rm eval}(K) = t_{0,\rm eval} + c_1 K$ ($t_{0,\rm eval}\approx 45\,\mu\mathrm{s}$, $c_1\approx 1.0\,\mu\mathrm{s}$), accurately captures the transition from the flat overhead floor to the asymptotic $K^2$ and $K$ scaling regimes out to $K=100$. Even at $K=100$, both operations remain several orders of magnitude faster than the $K$-independent $N\times N$ unbinned Cholesky setup ($1.58$~s) and triangular-solve evaluation ($38.4$~ms).

\section{Storage}
\label{sec:storage}

The inverse covariance matrix shares the low-rank structure of $\bC$: defining the $N\times K$ matrix $\bV\equiv \bD^{-1}\bU\bQ^{-1/2}$, Equation~\ref{eq:woodbury} takes the form $\bC^{-1}=\bD^{-1}-\bV\bV^T$. Storing either $(\bD,\bU)$ or the inverse representation requires only $N(K+1)$ numbers instead of $N^2$ (or $N(N+1)/2$ when packed), reducing the storage scaling from $O(N^2)$ to $O(NK)$ so that a precomputed $N\times N$ inverse file is never needed.

\begin{figure}[htbp]
\centering
\includegraphics[width=0.82\textwidth]{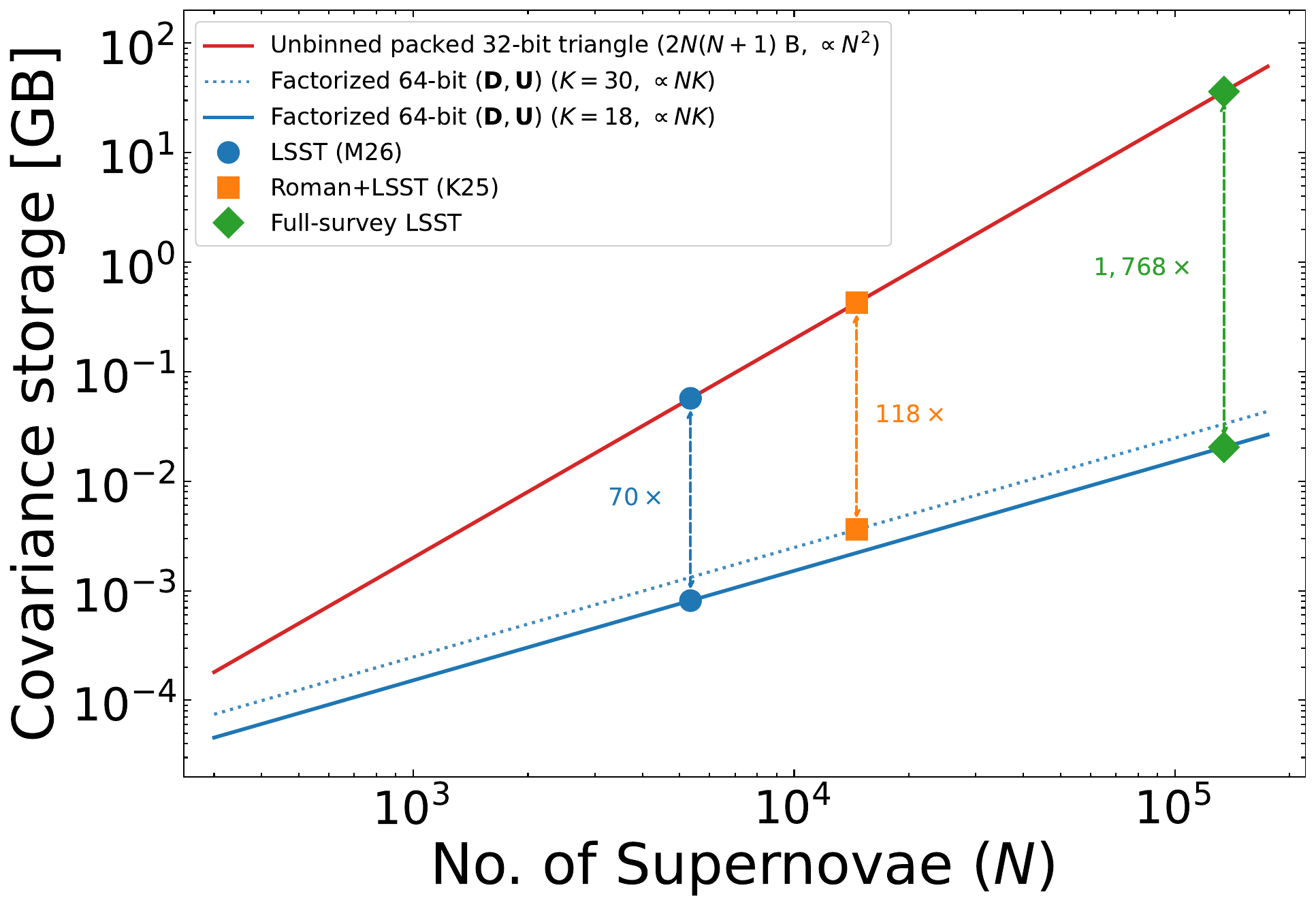}
\caption{Covariance storage as a function of the number of supernovae $N$, 
comparing 32-bit upper-triangular unbinned files 
($2N(N+1)$~bytes, solid red) 
against 64-bit factorized $(\bD,\bU)$ files ($8N(K+1)$~bytes for $K=18$ solid blue and $K=30$ dotted blue). 
Markers show the unbinned and factorized file sizes for LSST M26 
(circles; $57$~MB versus $0.8$~MB), 
Roman+LSST K25 (squares; $425$~MB versus $4$~MB), and 
10-yr LSST (diamonds; $36$~GB versus $20$~MB), 
with the corresponding storage compression factors annotated beside each vertical arrow.}
\label{fig:storage}
\end{figure}

As summarized in Table~\ref{tab:summary} and Figure~\ref{fig:storage}, 
the 64-bit factorized representation 
reduces on-disk covariance storage relative to a
32-bit upper-triangular unbinned file
by two to three orders of magnitude across current and upcoming survey scales: 
a factor of $70$ for M26 ($0.8$~MB versus $57$~MB), 
a factor of ${\sim}100$ for K25
($4$~MB versus $425$~MB), and 
a factor of $1{,}800$ for 10-yr LSST (20~MB vs. 36~GB).
Beyond a single covariance file, the unbinned quadratic storage growth poses a severe campaign-level bottleneck
for 10-yr LSST because analysis validation
routinely generates and archives covariance matrices across dozens of simulation realizations, 
and dozens of systematic options per realization.
For example,
225 $\covmat$
files were produced in the M26 analysis, and for 10-yr LSST
a single production campaign 
generating 225 $\covmat$ files
would require $8$~TB,
growing to tens of terabytes across iterative analysis passes.
In factorized form, the entire set of 225 $\covmat$ files would occupy
only 5~GB.

\section{Discussion}
\label{sec:discussion}

For systematic covariances built from a fixed set of distance-shift vectors, the computational trade-off between unbinned accuracy and binned speed identified by BHS21 and K23 is eliminated. Factorization  preserves the floating-point precision while reducing both likelihood evaluation and storage to $O(NK)$. This allows production pipelines to use the unbinned Hubble diagram not only for a single nominal fit, but also across the full suite of systematic-variant and sub-sample analyses where rebinning was previously used to reduce processing time.

The CPU reduction is not limited to the specific \SNANA\ cosmology fitting code that we tested.
The same $O(NK)$ likelihood evaluation and analytic normalization (Eq.~\ref{eq:marg_loglike}) 
can be used in Markov Chain Monte Carlo (MCMC) and nested-sampling frameworks such as 
{\tt CosmoSIS} \citep{Zuntz2015}, 
{\tt CosmoMC} \citep{LewisBridle2002}, and 
{\tt Cobaya} \citep{TorradoLewis2021}, 
where the overall inference speedup depends on the relative cost of the Boltzmann solver and external likelihoods. More broadly, the method applies to any
observational covariance that consists of independent per-object statistical noise plus a modest number of coherent systematic templates; analogous low-rank structures are exploited in pulsar-timing analyses \citep{VanHaasterenVallisneri2014} and multilevel factor models \citep{PHDB24,Parshakova2025}, whereas here the factorization follows as an exact algebraic identity from the definition of supernova systematic shift vectors.

\section{Conclusion}
\label{sec:conclusion}

The Binning is Sinning series established the importance of retaining individual-supernova information in cosmological fits, first through unbinned analyses (BHS21) and subsequently through redshift--stretch--color rebinning to keep large photometric samples computationally tractable (K23). By exploiting the exact low-rank structure $\bC=\bD+\bU\bU^T$ of systematic covariances built from $K$ per-supernova distance-shift vectors, the Woodbury identity reduces likelihood evaluation and storage from $O(N^2)$ (and $O(N^3)$ setup) to $O(NK)$ for $K\ll N$, removing the computational need for binning.

Implemented in \SNANA\ and validated across simulated 
LSST ($N\approx 5{,}300$, $K=18$), 
Roman+LSST ($N\approx 14{,}600$, $K=30$), and 
10-yr LSST ($N{\approx}134{,}400$, $K=18$) samples, 
the factorized method reproduces the unbinned best-fit cosmological parameters identically and matches uncertainties and the dark energy FoM to within $0.08\%$.
Compared to the traditional approach of using a dense covariance matrix,
single-step
likelihood evaluations are accelerated by a factor of $\sim 600$, 
$w_0w_a$CDM cosmology fits are accelerated by up to a factor of $63$ 
(completing in ${\sim}80$~minutes, even with $N>10^5$), 
and covariance disk storage is reduced by factors of 
$70$ to $1{,}800$ relative to unbinned files, 
keeping covariance files at the megabyte scale. 
Factorized unbinned inference therefore provides a practical, 
exact standard for next-generation supernova cosmology analyses.

\section*{Acknowledgments}
This research used resources of the National Energy Research Scientific Computing Center (NERSC), a Department of Energy User Facility supported by the Office of Science of the U.S. Department of Energy under Contract No.~DE-AC02-05CH11231, and was completed in part with resources provided by the University of Chicago Research Computing Center (RCC).
We acknowledge the use of AI agents (\href{https://shorturl.at/BniTV}{claude code} from Anthropic)  for \SNANA\ coding and manuscript preparation assistance; all  code, results, and plots were verified by the authors.   


\bibliographystyle{aasjournalv7}
\bibliography{references}

\end{document}